\documentstyle[preprint,eqsecnum,aps,epsfig]{revtex}
\tighten
\begin{document}
\draft
\preprint{UL--NTZ 31/98}
\title{
Production of para-- and orthopositronium at
relativistic heavy ion colliders 
}
\author{G. L. Kotkin\thanks{Email address: kotkin@math.nsc.ru}}
\address{Department of Physics,
         Novosibirsk State University, Novosibirsk, 630090, Russia}
\author{E. A. Kuraev\thanks{Email address: kuraev@thsun1.jinr.dubna.su}}
\address{Laboratory of Theoretical Physics,
  Joint Institute for Nuclear Research, Dubna, 141980 Russia}
\author{A. Schiller\thanks{Email address: schiller@tph204.physik.uni-leipzig.de}}
\address{Institut f\"ur Theoretische Physik and 
 Naturwissenschaftlich-Theoretisches Zentrum,
         Universit\"at Leipzig,  D-04109 Leipzig, Germany}
\author{V. G. Serbo\thanks{Email address: serbo@math.nsc.ru}}
\address{Department of Physics, 
         Novosibirsk State University, Novosibirsk, 630090, Russia}
\date{November 27, 1998}

\maketitle

\begin{abstract}
We consider the ortho-- and parapositronium production in the process 
$AA \to AA+$ Ps where A is a nucleus with the charge number Z. The inclusive 
cross section and the energy distribution of the relativistic Ps are calculated
which are of primary interest from the experimental point of view. The accuracy
of the corresponding cross sections is given by omitting terms 
$\sim (Z\alpha )^2/L^2$ for the para--Ps and $\sim (Z\alpha )^2/L$ for the 
ortho--Ps production where $L=\ln{\gamma^2} \approx 9$ and 16 for the RHIC and 
the LHC. Within this accuracy the multiphoton (Coulomb) corrections are taken 
into account. We show that the RHIC and the LHC will be Ps factories with a 
productions rate of about $10^5 \div 10^8$ relativistic Ps per day. The 
fraction of the ortho--Ps is expected to be of the same order as that of the 
para--Ps for Au--Au and Pb--Pb collisions.
\end{abstract}


\section{Introduction.}

Parapositronium (para--Ps, $n\, ^1\!S_0$) has a positive charge parity $C =+1$ 
and in the ground state, $n=1$, its life time at rest $\tau_0$ is 
$c\tau_0=3.7$ cm. On the other hand, orthopositronium (ortho--Ps, $n\, ^3\!S_1$)
has $C=-1$ and $c\tau_0=42$ m. The relativistic Ps has a decay length 
$c\tau_0\varepsilon/m_{\rm Ps}$ where $\varepsilon$ and 
$m_{\rm Ps}\approx 2m_e$ denote its energy and mass. The production of Ps in 
different collisions is of interest by the following reasons:
\begin{itemize}
\item
Ps can be used to test fundamental laws like the $CPT$-theorem (see review 
\cite{M}).
\item
A more detailed comparison between the experimental value of ortho--Ps width 
and its theoretical prediction is necessary since up to now there is an 
essential difference between them \cite{KhM}).
\item
Relativistic Ps can be used to study the unusual large transparency of Ps in 
thin layers predicted in Refs.~\cite{N,LP}.
\end{itemize}

Up to now the following methods have been proposed to create positronium:
The photo-\cite{MSS}-\cite{TC} and electroproduction \cite{MSS,ABN,HO} of Ps 
on nuclei and atoms. The results of Refs.~\cite{MSS}-\cite{HO} were completed 
and corrected in Ref.~\cite{GKSST}. Recently a new method, based on an electron
cooling technique, was proposed in the works \cite{M,MS}. It was claimed that 
this method can give about $10^4$ nonrelativistic Ps per second on a 
specifically designed accelerator.

In the present paper we study in detail a new and promising possibility --- the 
production of relativistic Ps at colliders with relativistic heavy nuclei like 
the RHIC and the LHC, i.e. we consider the process $AA\to AA+{\rm Ps}$ where 
$A$ denotes the nucleus with charge number $Z$. The parameters of these 
colliders (from Refs.~\cite{PDG}) are given in Table~\ref{tab1}. We show that 
at these colliders
\begin{equation}
\sim 1 \div 10^3 \mbox{ relativistic Ps per second }
\label{1}
\end{equation}
will be produced.It is  interesting to note that the ratio of ortho--Ps to 
para--Ps events turns out to be large: for the RHIC about 0.8, for the LHC 0.3 
in the Pb--Pb mode and 0.02 in the Ca--Ca mode.

\begin{figure}[!htb]
  \begin{center}
    \epsfig{file=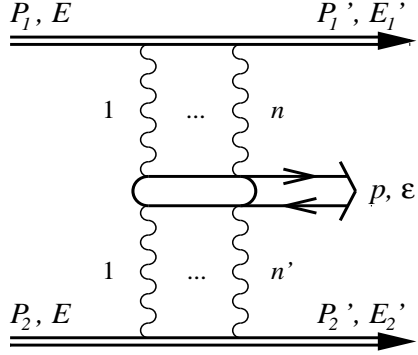,width=46mm,height=55mm,angle=270}
    \vspace{3mm} 
    \caption{The amplitude $M_{nn'}$ for the production of Ps by $n\;(n')$
             virtual photons emitted by the first (second) nucleus.}
    \label{fig1}
  \end{center}
\end{figure}

Up to now only the total cross section for para--Ps production has been 
calculated in the  leading logarithmic approximation (LLA) 
(see Ref.~\cite{MSS}). There it has been found:
\begin{equation}
\sigma^{\mathrm{LLA}}_{\mathrm{para}}={1\over 3}\, \zeta (3)\,\sigma_0\,
L^3\,,\quad \sigma_0={Z^4\alpha^7\over m^2_e},\quad \zeta
(3)=\sum^\infty_{n=1}{1\over n^3} =1.202 \dots \,,\quad
L=\ln \gamma^2 \,.
\label{2}
\end{equation}
However, for experiments (especially with relativistic Ps), it is much more 
interesting to know the energy and angular distributions of Ps which are given 
by the inclusive cross section and energy spectrum of Ps. These distributions 
for para--Ps as well as for ortho--Ps are calculated in the present work. In our 
calculations we also take into account with a certain accuracy the Coulomb 
correction (CC) related to multiphoton exchange of the produced Ps with nuclei 
described by Feynman diagrams of Fig.~\ref{fig1}. They are quite important and 
decrease the cross sections by $30 \div 50$ \% for the Au--Au and Pb--Pb 
collisions. The approach used here is based essentially on results on the 
production of relativistic positronium in collisions of photons and electrons 
with heavy ions presented in our recent work~\cite{GKSST}.

Our main notations are  given in Eq.~(\ref{2}) and Fig.~\ref{fig1}: the 
{\it i}-th nucleus changes its 4-momentum (energy) $P_i\;(E)$ to 
$P'_i\; (E'_i)$ and emits $n$ or $n'$ virtual photons with the total 
4-momentum $q_i=P_i-P'_i$ and the energy $\omega_i=E-E'_i$. These photons 
collide and produce Ps with 4-momentum $p=q_1+q_2$ and energy 
$\varepsilon=\omega_1+\omega_2$. Due to charge parity conservation, the sum 
$n+n'$ is even (odd) for para--Ps (ortho--Ps) production. We use the c.m.s. with 
the $z$--axis along the vector ${\bf P}_1$ and denote by ${\bf p}_\perp$ the 
transverse components of vector ${\bf p}$. The CC depends on the parameter
\begin{equation}
\nu=Z\alpha \approx Z/137\,.
\label{3}
\end{equation}

In Section 2 we discuss the production of para--Ps in ultrarelativistic heavy 
ion collisions. After selecting the leading diagrams including the multiphoton 
exchange with the nuclei, the Born cross sections are calculated in the 
equivalent photon approximation to identify the dominant kinematical regions. 
The exact Born cross sections for ultrarelativistic ions are found in the next 
subsection. The CC are calculated in the accuracy given by the selection of the
leading diagrams. Section 3 is devoted to the production of ortho--Ps. Our 
results are summarized in Section 4 where we discuss the expected production 
rates for the colliders RHIC and LHC.

\section{Production of para--Ps}

\subsection{Selection of the leading diagrams including the multiphoton 
exchange}
\label{select}
The amplitude $M_{\mathrm{para}}$ of the para--Ps production can be
presented in the following form via the sums over amplitudes
$M_{nn'}$ of Fig.~1 (with $n+n'$ being even)
\begin{equation}
M_{\mathrm{para}}= \sum_{n,n' \geq 1}\, M_{nn'}=
M_{\rm B}+M_1+\tilde M_1+M_2 \,,\quad
M_{\rm B} \equiv M_{11}\,,
\label{4}
\end{equation}
$$
M_1=\sum_{n'=3,5,7,...} M_{1n'}\,, \quad
\tilde M_1=\sum_{n=3,5,7,...}M_{n1}\,,\quad
M_2=\sum_{n,n'\geq 2}M_{nn'}\,.
$$
Here $M_{\rm B}$ is the Born amplitude in which each nucleus emits one photon, 
the item $M_1$ ($\tilde M_1)$ corresponds to the case when the first (second) 
nucleus emits one photon while the other nucleus emits at least three photons, 
the last term $M_2$ corresponds to the case when both nuclei emit at least two 
photons. Using this decomposition of the amplitude, we present the cross 
section in the form
$$
\sigma_{\rm para}=\sigma_{\rm B}+
\sigma_1+\tilde\sigma_1+\sigma_2\,, \quad
d\sigma_{\rm B} \propto \, |M_{\rm B}|^2\,,
$$
\begin{equation}
d\sigma_1 \propto\, 2{\rm Re}\,(M_{\rm B} M^*_1)+|M_1|^2 \,,\;\;
d\tilde\sigma_1 \propto\, 2{\rm Re}\,(M_{\rm B} \tilde M^*_1)+|\tilde
M_1|^2 \,,
\label{5}
\end{equation}
$$
d\sigma_2 \propto\, 2{\rm Re}\,(M_{\rm B} M^*_2+ M_1\tilde M^*_1 +
M_1 M^*_2 + \tilde M_1 M^*_2) +|M_2|^2 \,.
$$

It is not difficult to estimate that the integration over the total virtual 
photon momenta squared $q^2_1$ and $q^2_2$ gives two large 
Weizs\"acker--Williams (WW) logarithms $\sim L^2$ in $\sigma_{\rm B}$, one 
large WW logarithm $\sim L$ in $\sigma_1$ and $\tilde \sigma_1$ and no one 
large WW logarithm in $\sigma_2$. Therefore, the relative contributions of the 
items $\sigma_i$ is 
$\sigma_1/\sigma_{\rm B}=\tilde\sigma_1/\sigma_{\rm B} \sim \nu^2/L$ and
\begin{equation}
{\sigma_2 \over\sigma_{\rm B}}\, \sim \, {\nu^2\over L^2}\, <\,
0.4 \%
\label{6}
\end{equation}
for the colliders in Table~\ref{tab1}. As a result, with
an accuracy of the order of 1 \% we can neglect $\sigma_2$ and
use the equation
\begin{equation}
\sigma=\sigma_{\rm B}+\sigma_1+\tilde \sigma_1\,.
\label{7}
\end{equation}
Moreover, with the same accuracy we can calculate $\sigma_1$ and
$\tilde\sigma_1$ in LLA only since the terms next to the leading ones
are of the order of $\nu^2/L^2$.

\subsection{Calculating the Born cross section in the equivalent
photon approximation}
\label{EPA}
Before we proceed to a more exact calculation of $d\sigma_{\rm B}$,
it is useful to perform a rough estimate using the equivalent
photon approximation (EPA). This allows us to
identify the dominant region in phase space.

It is well known that the main contribution to the pair production 
$AA\to AA e^+ e^-$ is given by the region of small photon virtualities 
$-q^2_i \ll (q_1+q_2)^2$. Therefore, for the Ps production this region 
is\footnote{This can also be seen from the expression for the Born amplitude 
obtained in Appendix A.}
\begin{equation}
-q^2_i \ll 4m^2_e\,,
\label{8}
\end{equation}
i.e. the virtual photons are almost real in the dominant region. Since
\begin{equation}
-q^2_i={\bf q}^2_{i\perp}+(\omega_i/\gamma)^2\,,
\label{9}
\end{equation}
we immediately conclude that in the dominant region the photon
energies $\omega_i$ are small compared with the nucleus energies $E$:
\begin{equation}
\omega_i \stackrel{<}{\sim} m_e\gamma \ll E\,.
\label{10}
\end{equation}

In the EPA the cross section for the $AA\to AA+{\rm Ps}$ process is related to 
the cross section for the real photoprocess $\gamma\gamma\to {\rm Ps}$ by
\begin{equation}
d\sigma_{\rm B}= dn_1(\omega_1)\, dn_2(\omega_2)\,
\sigma_{\gamma\gamma\to {\rm Ps}} (\omega_1\omega_2)\,,\quad
\omega_{1,2}\approx {1\over 2}(\varepsilon \pm p_z)\,,
\label{11}
\end{equation}
where $dn_i$ is the number of the equivalent photons  generated
by the {\it i}-th nucleus. For $dn_i$ we use two equivalent forms
(see, for example, formulae from Appendix D of review~\cite{BGMS}
in which we omit terms of the order of $\omega_i/E$ (in virtue of
relation~(\ref{10}))
\begin{equation}
dn_i={Z^2\alpha \over \pi^2} \, {{\bf q}^2_{i\perp} \over
(q^2_i)^2} \,{d\omega_i\over \omega_i}\, d^2 q_{i\perp}\,,
\label{12}
\end{equation}
\begin{equation}
dn_i={Z^2\alpha \over\pi} \, \left[ 1-{(\omega_i/\gamma)^2\over
(-q^2_i)}\right]\, {d\omega_i\over\omega_i}\, {d(-q^2_i)
\over(-q^2_i)}\,{d\varphi_i\over 2\pi}\,,
\label{13}
\end{equation}
where $\varphi_i$ is the  azimuthal angle of the vector ${\bf
q}_i$. It is also well known that the  equivalent photons of small
energy are completely linearly polarized in the scattering plane
and their polarization vectors are
\begin{equation}
{\bf e}_i={{\bf q}_{i\perp}\over|{\bf q}_{i\perp}|}\,.
\label{14}
\end{equation}

The cross section for the real photoprocess $\gamma\gamma\to
{\rm Ps}$ can be easily obtained using the known width of the
$n\,^1\!S_0$ state $\Gamma_{{\rm Ps}\to\gamma\gamma}=\alpha^5 m_e
\, ({\bf e}_1 \times {\bf e}_2)^2 / (4n^3)$ and summing over
all para--Ps states $n$
\begin{equation}
\sigma_{\gamma \gamma\to {\rm Ps}}= 4\pi^2\alpha^5\zeta (3)
({\bf e}_1 \times {\bf e}_2)^2 \delta(4\omega_1\omega_2-4m^2_e)\,.
\label{15}
\end{equation}
Substituting Eqs.~(\ref{13})-(\ref{15}) into (\ref{11}) and
integrating over $\varphi_i,\; \omega_2$ and over $q^2_i$ in the
region
\begin{equation}
(\omega_i/\gamma)^2 \stackrel{<}{\sim} -q^2_i \stackrel{<}{\sim}
m^2_e
\label{16}
\end{equation}
we find with logarithmic accuracy
\begin{equation}
d\sigma_{\rm B}=2\zeta(3) \sigma_0 \,\ln{m_e\gamma\over\omega_1}\,
\ln{m_e\gamma\over\omega_2} \,{d\omega_1\over\omega_1}\,,\quad
\omega_1\omega_2 \approx m^2_e \,.
\label{17}
\end{equation}
Taking into account the inequalities (\ref{10}), we conclude that
the main region for $\omega_1$ is
\begin{equation}
{m_e\over\gamma}  \stackrel{<}{\sim} \omega_1 \stackrel{<}{\sim}
m_e\gamma \,.
\label{18}
\end{equation}
Integrating $d\sigma_{\rm B}$ (\ref{17})  over the photon energy $\omega_1$
in the region (\ref{18}) we obtain the total cross  section (\ref{2}).

The {\bf spectrum} of the relativistic Ps can be obtained from Eq. (\ref{17}) 
as follows. Let us divide the region (\ref{18}) into two symmetrical 
subregions. In the first one, $m_e \ll \omega_1 \stackrel{<}{\sim} m_e\gamma$, 
we have $\omega_1 \approx \varepsilon \approx p_z$, 
$\omega_2 \approx m^2_e/\varepsilon$ and the contribution to the spectrum is 
\begin{equation}
d\sigma^{(1)}_{\rm B}=2\zeta(3)\, \sigma_0 \,
\ln{m_e\gamma\over \varepsilon} \,
\ln{\varepsilon\gamma\over m_e} \,{d\varepsilon\over\varepsilon} \,.
\label{19}
\end{equation}
In the second subregion $m_e/\gamma \stackrel{<}{\sim}
\omega_1 \ll m_e$ we have $\omega_2 \approx \varepsilon
\approx -p_z$, $\omega_1 \approx m^2_e/\varepsilon$ and
\begin{equation}
d\sigma^{(2)}_{\rm B}=2\zeta(3) \, \sigma_0 \,
\ln{\varepsilon\gamma\over m_e}\, \ln{m_e\gamma \over
\varepsilon} \, {d\varepsilon\over\varepsilon} \,.
\label{20}
\end{equation}
The total contribution of these two subregions is
\begin{equation}
d\sigma_{\rm B}=4\zeta(3) \,\sigma_0 \,
\ln{m_e\gamma\over\varepsilon}\, \ln{\varepsilon\gamma\over
m_e}\,{d\varepsilon\over\varepsilon}\,, \;\; m_e \ll \varepsilon
\stackrel{<}{\sim} m_e\gamma \,.
\label{21}
\end{equation}
Integrating this spectrum over $\varepsilon$ leads to the result (\ref{2}).

To calculate the {\bf inclusive cross section} we substitute Eqs.~(\ref{12}), 
(\ref{14}), (\ref{15}) into (\ref{11}) and perform a simple integration
$$
\int \,{d\omega_1\over\omega_1} \, {d\omega_2\over\omega_2}\,
\delta(4\omega_1\omega_2-4m^2_e)={1\over 4m^2_e}\, {d
p_z\over\varepsilon}\,.
$$
Further, we insert into  the right-hand side of Eq.~(\ref{11}) the factor
$$
 1=\int \, \delta({\bf q}_{1\perp}+{\bf q}_{2\perp}-{\bf
p}_\perp ) \, d^2 p_\perp
$$
and find the  inclusive cross section in the form
\begin{equation}
\varepsilon \, {d^3\sigma_{\rm B}\over d^3 p}=
{\zeta(3)\over\pi^2} \, \sigma_0 \, \int \, {({\bf q}_{1\perp}
\times {\bf q}_{2\perp})^2 \over (q^2_1\, q^2_2)^2} \,
\delta({\bf q}_{1\perp}+ {\bf q}_{2\perp}-{\bf p}_\perp ) \,
d^2 q_{1\perp} d^2 q_{2\perp}\,.
\label{22}
\end{equation}
{}From this expression and taking into account Eqs.~(\ref{9}),
(\ref{16}), we conclude that the leading logarithmic contribution
to the cross section is given by the region
\begin{equation}
\left({\omega_1\over\gamma}\right)^2+
\left({\omega_2\over\gamma}\right)^2 \ll {\bf p}^2_\perp \ll m^2_e \,.
\label{23}
\end{equation}
It arises from the two symmetrical subregions:
\begin{eqnarray}
&&\left({\omega_1\over\gamma}\right)^2 \ll {\bf q}^2_{1\perp} \ll
{\bf p}^2_\perp \,,\quad {\bf q}_{2\perp} \approx {\bf p}_\perp \,,
\nonumber \\
&&\left({\omega_2\over\gamma}\right)^2 \ll {\bf q}^2_{2\perp} \ll
{\bf p}^2_\perp \,,\quad {\bf q}_{1\perp} \approx {\bf p}_\perp \,.
\label{24}
\end{eqnarray}
The contribution from the first region is
$$
{\zeta(3)\over 2\pi^2} \, \sigma_0 \,
\int_0^{2\pi} d\varphi \,
 \int_{(\omega_1/\gamma)^2}^{p^2_\perp} d {\bf q}^2_{1\perp}\,
\, {({\bf q}_{1\perp}
\times {\bf p}_{\perp})^2 \over ({\bf q}^2_{1\perp} {\bf p}^2_\perp)^2} \,
  =
{\zeta(3)\over 2\pi} \,{\sigma_0\over {\bf p}^2_\perp} \,
\ln{{\bf p}^2_\perp \gamma^2 \over \omega_1^2} \,.
$$
Taking into account the second region, we find the whole inclusive cross 
section
\begin{equation}
\varepsilon \,{d^3\sigma_{\rm B}\over d^3 p}=
{\zeta(3)\over\pi} \,{\sigma_0\over {\bf p}^2_\perp} \,
\ln{{\bf p}^2_\perp \gamma^2 \over m^2_e} \,.
\label{25}
\end{equation}
Integrating this expression over ${\bf p}^2_\perp$ in the range (\ref{23}) and 
taking into account the two symmetrical subregions (\ref{19},\ref{20}) 
mentioned before, we obtain the spectrum (\ref{21}).

\subsection{The exact calculation of the Born cross section for 
ultrarelativistic ions}

The Born amplitude of the discussed process is  obtained in
Appendix A omitting terms of the relative order of $1/\gamma^2$,
$m_e/M$ or smaller. Using this expression for the amplitude we can
present the {\bf inclusive cross section} in the form
\begin{equation}
\varepsilon \, {d^3\sigma_{\rm B}\over d^3 p}= {\zeta(3)\over
4\pi}\, {\sigma_0\over m^2_e}\, J_{\mathrm{B}}\,,
\label{26}
\end{equation}
$$
 J_{\mathrm{B}}={4m^2_e \over
\pi} \, \int \, {\bf A}^2 \, \delta({\bf q}_{1\perp}+{\bf
q}_{2\perp}- {\bf p}_\perp ) \, d^2 q_{1\perp} d^2 q_{2\perp} \,,
$$
$$
{\bf A}={{\bf q}_{1\perp}\times{\bf q}_{2\perp}\over q^2_1\,
q^2_2}\; {4m^2_e\over 4m^2_e-q^2_1-q^2_2}
$$
which differs from the approximate Eq.~(\ref{22}) only by the
factor $(4m^2_e)^2/(4m^2_e-q^2_1-q^2_2)^2$ in the integrand. It
is convenient to introduce the  dimensionless variables
\begin{equation}
x={\varepsilon+p_z\over 4m_e\gamma}\,,\quad
\tilde x={\varepsilon-p_z\over 4m_e\gamma}\,,\quad
\tau=\left( {{\bf p}_\perp\over 2m_e}\right)^2 \,.
\label{27}
\end{equation}
They are not independent, but connected by the constraint
\begin{equation}
x \tilde x={1+\tau\over 4\gamma^2}\,.
\label{28}
\end{equation}
In this notation
\begin{equation}
-q^2_1={\bf q}^2_{1\perp}+(2m_e x)^2 \,,\quad
-q^2_2={\bf q}^2_{2\perp}+(2m_e\tilde x)^2 \,.
\label{29}
\end{equation}

It is useful to note that the rapidity of the Ps is equal to 
$(1/2)\, \ln(x/\tilde x)$ and that the variables $x$ and $\tilde x$ have a 
very simple form for relativistic Ps. Indeed, for $p_z \gg m_e$ the variable 
$x$ is $x=\varepsilon/(2m_e\gamma)$, i.e. $x$ is the ratio of Ps energy to the 
characteristic energy $2m_e\gamma$ (above $2m_e\gamma$ the spectrum  drops 
very quickly). Analogously, for $(-p_z) \gg m_e$ 
$\tilde x=\varepsilon /(2m_e\gamma)$. The neglected contributions are of the 
order of $(4m_e^2 + {\bf p}^2_\perp)/ \varepsilon^2$.
{}From the discussion in 
Sect.~\ref{EPA} it follows that the dominant contribution to the cross i
section (\ref{26}) is given by the  region
\begin{equation}
{1\over\gamma^2} \stackrel{<}{\sim} x \,,\, \tilde x
\stackrel{<}{\sim} 1 \,,
\quad x^2+\tilde x^2 \stackrel{<}{\sim} \tau
\stackrel{<}{\sim} 1 \,.
\label{30}
\end{equation}
In Appendix B we show that the function $J_{\rm B}$ can be
reduced to the one--dimensional integral
\begin{equation}
J_{\rm B}={\tau\over 8} \, \int^1_0 \, u^3 (1-u) (2-u)^2
K(u,x,\tau)\, du\,,
\label{31}
\end{equation}
\begin{eqnarray}
&K(u,x,\tau)&=3 \int^1_{-1} \, {(1-t^2)dt \over (at^2+bt+c)^4}
\label{32}
\\
&&={12a(b^2-5a^2+ac)\over d^{7/2}} \, \ln{c-a+\sqrt d \over c-a -\sqrt d}+
{4 N \over [(a+c)^2-b^2]^2 d^3} \,, 
\nonumber
\end{eqnarray}
\begin{eqnarray*}
N&=& -30a^6+b^6-104 a^5 c+56 a^4 b^2-124 a^4 c^2+ 90 a^3 b^2 c-56 a^3 c^3
\\
&&-26 a^2 b^4+36 a^2 b^2 c^2-6a^2 c^4-2ab^4 c-6 ab^2 c^3
\end{eqnarray*}
where
$$
a=-{1\over 4} u^2\tau\,,\quad b={1\over 2} \, u (2-u) \,(x^2-\tilde x^2)\,,
$$
\begin{equation}
c={1\over 4}(2-u)^2\tau+{1\over 2}(2-u)^2(x^2+\tilde x^2)+
(1-u)(2-u)\,,\quad d=b^2-4ac\,.
\label{33}
\end{equation}
Let us stress that the inclusive cross section (\ref{26})
with (\ref{31}) is valid both for relativistic para--Ps with
$\varepsilon \gg 2m_e$ as well and for nonrelativistic para--Ps
with $\varepsilon \sim 2 m_e$.

Using the representation (\ref{26}) for $J_{\mathrm{B}}$ it is not difficult 
to find that the inclusive cross section is large in the main region
\begin{equation}
J_{\rm B}={1\over \tau} \left[\ln(4\gamma^2\tau)-1\right] \;\;
\mbox{ at } \;\; x^2+\tilde x^2 \ll \tau \ll 1
\label{34}
\end{equation}
but drops very quickly outside this region. Indeed, the function
$J_{\rm B}$ is small at large $\tau$
\begin{equation}
J_{\rm B}={1\over \tau^3} \, \ln(4\gamma^2) \;\; \mbox{ at }
\;\; x^2+\tilde x^2 \ll 1 \ll \tau\,,
\label{35}
\end{equation}
at large $x$
\begin{equation}
J_{\rm B}={\tau\over x^8} \, \ln(4\gamma^2 x^2) \;\; \mbox{ at
}\;\; \tau \ll 1 \ll x^2\,,
\label{36}
\end{equation}
and at very small $\tau$
\begin{equation}
J_{\rm B}={\tau\over (x^2-\tilde x^2)^2} \, \left( {x^2+\tilde
x^2\over x^2-\tilde x^2} \, \ln{x \over \tilde x } - 1 \right)
\;\; \mbox{ at } \;\; \tau \ll x^2+\tilde x^2 \ll 1\,.
\label{37}
\end{equation}

\begin{figure}[!htb]
  \begin{center}
   \epsfig{file=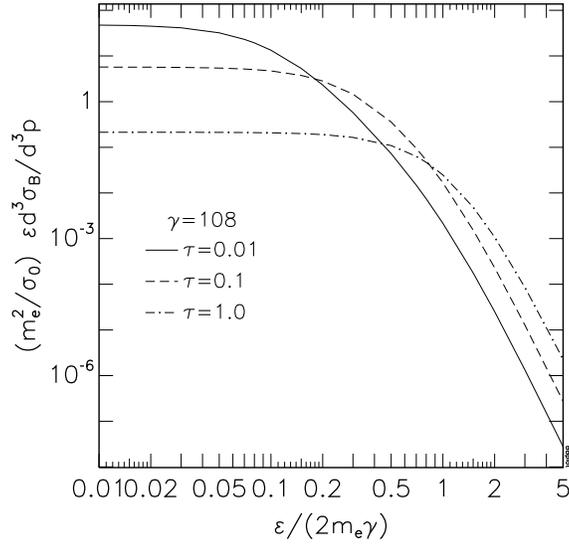,width=75mm}
   \caption{The inclusive cross section (\ref{26}) as function of
           the scaled para--Ps energy $\varepsilon/(2 m_e \gamma)$ 
           for the RHIC at different values of $\tau =
          {\bf p}^2_\perp / (2m_e)^2$.}
   \label{fig2}
  \end{center}
\end{figure}

\begin{figure}[!htb]
  \begin{center}
    \epsfig{file=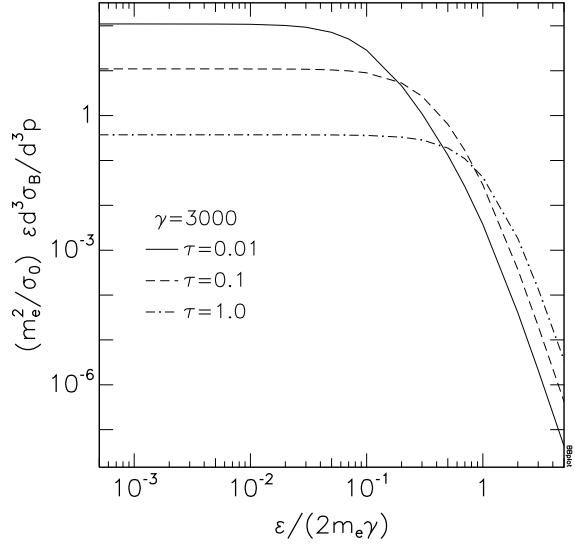,width=75mm}
     \caption{Same as Fig.~\protect\ref{fig2} for the LHC (Pb--Pb mode)}
     \label{fig3}
   \end{center}
\end{figure}

In Figs.~\ref{fig2},\ref{fig3} we present the inclusive cross section 
(\ref{26}) at the RHIC and the LHC (Pb--Pb mode) as function of the scaled Ps 
energy for different scaled transverse momenta squared  $\tau$. 
{}From these Figures it is obvious that for 
$\varepsilon/(2 m_e \gamma)<\sqrt{\tau}$ the 
cross sections only weakly depend on the Ps energy. The cross sections quickly 
drop at larger energies.

\begin{figure}[!htb]
  \begin{center}
     \epsfig{file=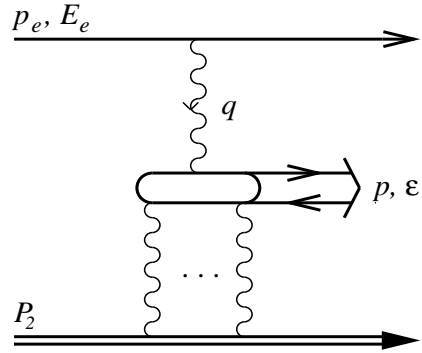,width=46mm,height=55mm,angle=270}
     \vspace{3mm}
     \caption{The electroproduction of Ps on a nucleus.}
     \label{fig4}
  \end{center}
\end{figure}

To obtain the {\bf spectrum} of relativistic Ps it is convenient to use an 
approach developed in Ref.~\cite{GKSST} for the electroproduction of Ps on 
nucleus $eA\to eA+{\rm Ps}$ (Fig.~\ref{fig4}). It is well known that the 
electroproduction cross section can be exactly written in terms of two 
structure functions or two cross section $\sigma_T (s_\gamma,q^2)$ and 
$\sigma_S(s_\gamma,q^2)$ for the virtual processes $\gamma^*_T A\to {\rm Ps}+A$
and $\gamma^*_S A \to {\rm Ps}+A$ (where $\gamma^*_T$ and $\gamma^*_S$ denote 
the transverse and scalar virtual photons with 4-momentum $q$ and helicity 
$\lambda_T=\pm 1$ and $\lambda_S=0$, respectively, and $s_\gamma=2qP_2$):
\begin{equation}
d\sigma(eA\to {\rm Ps}+A)=\sigma_T(s_\gamma,q^2) \,
dn_T(s_\gamma,q^2) +\sigma_S(s_\gamma,q^2) \,dn_S(s_\gamma,q^2)\,.
\label{38}
\end{equation}
Here the coefficients $dn_T$ and $dn_S$ can be called the number
of the transverse and scalar photons generated by the electron.

To calculate  $d\sigma_{\rm B}$ in our case we can use a similar expression 
where the number of photons are now generated from a scalar projectile nucleus.i
The cross sections $\sigma_T$ and $\sigma_S$ for para--Ps have been obtained in 
Ref.~\cite{GKSST} for high energetic virtual photons (for $s_\gamma\gg 4m_e M$ 
or $\gamma^2 x \gg 1$):
\begin{equation}
\sigma_T=\pi\zeta(3) \, {Z^2\alpha^6\over m^2_e} \,
{L(x,y)-1\over (1+y)^2}\,, \quad \sigma_S=0
\label{39}
\end{equation}
with
\begin{equation}
L(x,y)=\ln(4\gamma^2 x)-{1\over 2}\ln(1+y)\,,\quad 
y={-q^2_1\over 4m^2_e} \,.
\label{40}
\end{equation}
Here $x$ is defined in Eq.~(\ref{27}). The accuracy of Eq.~(\ref{39})
is determined by neglecting terms $\sim 1/(\gamma^2 x)$,
$\sqrt{-q^2_1}/(m_e\gamma^2 x)$. The number of photons  generated
by the first nucleus are (see, for example Appendix D from Ref.
\cite{BGMS})
$$
dn_T={Z^2\alpha\over\pi} \,\left(1-{x^2\over y}\right) \,
{dx\over x}\, {dy\over y}\,,
$$
\begin{equation}
dn_S={Z^2\alpha\over\pi} \,{dx\over x} \,{dy\over y}\,.
\label{41}
\end{equation}
Note that $dn_T$ corresponds to $dn_1$ as given in Eq.~(\ref{13}).

Using Eqs.~(\ref{39})-(\ref{41}) and integrating over $y$ we 
obtain\footnote{The lower limit is $y_{min}=x^2$, the upper limit can be 
extended to infinity due to the fast convergence of the integral.}
\begin{equation}
d\sigma_{\rm B}=\zeta(3)\, \sigma_0 \, \left\{ f_1(x) \,
\left[ \ln(4\gamma^2 x) -1 \right] - f_2(x) \right\} \,{dx\over x}
\label{42}
\end{equation}
with
\begin{eqnarray}
&f_1(x)&=\int^\infty_{x^2} \, \left(1-{x^2\over y}\right) \,
{dy\over y(1+y)^2} = \left(1+2 x^2\right) \, \ln{1+x^2\over
x^2}\,-\,2 \,,
\nonumber \\ 
&f_2(x)&={1\over 2}\, \int^\infty_{x^2} \, \left(1-{x^2\over y}
\right) \,{\ln(1+y) \over y}\, {dy\over (1+y)^2} 
\label{43} \\
&&= \frac{1+2 x^2}{2} \, \left[ \frac{\pi^2}{6}
-{\mathrm {Li}}_2 \left( \frac{x^2}{1+x^2} \right)
 \right] - {1\over 2} x^2
\, \ln{1+x^2\over x^2}-{1\over 2}-\ln(1+x^2) \,.
\nonumber
\end{eqnarray}
Here the  dilogarithmic function is defined as
\begin{equation}
{\rm Li}_2(x)=-\int^x_0 \,{\ln(1-t)\over t} \,dt \,.
\label{44}
\end{equation}
For small $x\ll 1$ we have
\begin{equation}
f_1(x)= 2 \left( \ln{1\over x}-1 \right)\,,\quad
f_2(0)={\pi^2-6\over 12}=0.3225
\label{45}
\end{equation}
while for  large $x\gg 1$
\begin{equation}
f_1(x)={1\over 6x^4}\,,\quad f_2(x)={1\over 6x^4} \, \left(\ln x
+{5\over 12} \right) \,.
\label{46}
\end{equation}

In obtaining Eq.~(\ref{42}), we have considered the first nucleus as projectile
and the second nucleus as target. Taking the first nucleus as target and the 
second nucleus as  projectile we find for $d\sigma_{\rm B}$ the same 
expression as in Eq.~(\ref{42}) with the replacement $\; x \to \tilde x$:
\begin{equation}
d\sigma_{\rm B}=\zeta(3)\, \sigma_0 \, \left\{ f_1(\tilde x) \,
\left[ \ln(4\gamma^2 \tilde x) -1 \right] - f_2(\tilde x) \right\} \,
{d\tilde x\over \tilde x}\,.
\label{42b}
\end{equation}

To obtain the spectrum $d\sigma_{\rm B}/d\varepsilon$ we have to take into 
account the following two circumstances for the relativistic Ps: i
({\it i}) the variables $x$ and $\tilde x$ are related in a simple way to 
$\varepsilon$; ({\it ii}) there are two symmetrical regions in the c.m.s.: 
$p_z \approx \varepsilon \gg 2m_e$ and $(-p_z) \approx \varepsilon \gg 2m_e$. 
In the first region the variable $x$ is $x=\varepsilon/(2m_e\gamma)$ and we 
have to use Eq.~(\ref{42}), in the second region 
$\tilde x=\varepsilon/(2m_e\gamma)$ where Eq.~(\ref{42b}) is used. As a result,
the spectrum for the relativistic Ps has the form (compare with the 
approximate Eq.~(\ref{21}))
\begin{equation}
d\sigma_{\rm B}=2\zeta(3) \, \sigma_0 \, \left[ f_1\left(
\varepsilon\over 2m_e\gamma \right) \,
\left(\ln{2\varepsilon\gamma\over m_e} \,-\, 1 \right)
-f_2\left({\varepsilon\over 2m_e\gamma}\right)\right]\,
{d\varepsilon\over\varepsilon} \,.
\label{47}
\end{equation}
The accuracy of this expression is determined by the omitted
terms of the order of $1/\gamma^2$, $2m_e/ (\varepsilon \gamma)$,
$(2m_e/\varepsilon)^2$.

Note that in the main region $2m_e \ll \varepsilon \ll 2m_e \gamma$
the spectrum is logarithmically enhanced
\begin{equation}
d\sigma_{\rm B}=4\zeta(3) \, \sigma_0 \, \left[
\ln{2m_e \gamma \over \varepsilon} \,
\ln{2\varepsilon \gamma\over m_e} - \ln(4\gamma^2) +\frac{30-\pi^2}{24}
\right] \,
{d\varepsilon\over\varepsilon}
\label{47b}
\end{equation}
while for large $\varepsilon \gg 2m_e \gamma$ the spectrum drops very quickly
\begin{equation}
d\sigma_{\rm B}= {\zeta(3) \over 3} \, \sigma_0 \,
\left(\frac{2m_e \gamma}{\varepsilon}\right)^4 \,
\left[
\ln(4\gamma^2) - {17\over 12}
\right] \,
{d\varepsilon\over\varepsilon} \,.
\label{47c}
\end{equation}

\begin{figure}[!htb]
  \begin{center}
     \epsfig{file=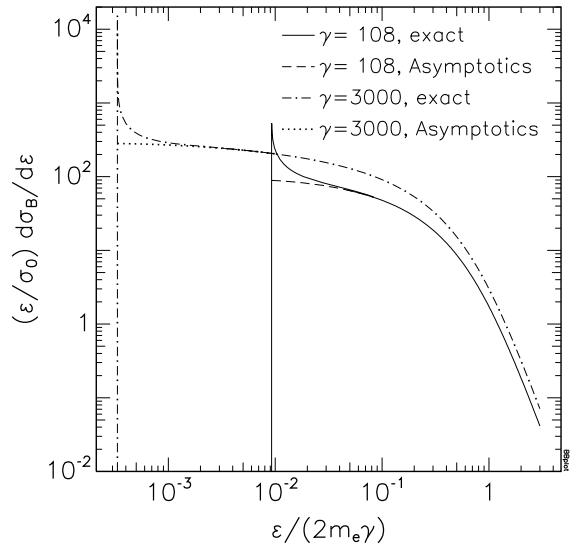,width=75mm}
     \caption{The spectrum of the para--Ps as function of
              the scaled Ps energy $\varepsilon/(2 m_e \gamma)$.}
     \label{fig5}
  \end{center}
\end{figure}

The spectrum of the para--Ps is shown in Fig.~\ref{fig5}. The solid and 
dash-dotted lines present the result for the RHIC and the LHC (Pb--Pb mode) 
obtained by a numerical integration of the inclusive cross section 
(\ref{26}), (\ref{31}). The dashed and dotted lines present the spectrum for 
both colliders calculated from Eq. (\ref{47}) which is valid for relativistic 
Ps with $\varepsilon \gg 2m_e$. It can be seen that in the region 
$10 m_e < \varepsilon < (0.1 \div 0.2) \,2m_e\gamma $ the quantity 
$\varepsilon d\sigma_{\rm B}/ d \varepsilon$ has a very weak dependence on 
$\varepsilon$ whereas the spectra drop quickly for 
$\varepsilon /( 2 m_e \gamma) \stackrel{>}{\sim} 0.5$.

{}From the experimental point of view, it is interesting to integrate the 
cross section  over the region $\varepsilon \geq \varepsilon_{\min}$
(where $ 2m_e \ll\varepsilon_{\min} \ll 2m_e \gamma$).
It can be obtained from Eq.~(\ref{47}) omitting pieces
of the order of $\varepsilon_{\min}^2 / [(2m_e \gamma)^2 \, l]$:
\begin{equation}
\sigma_{\rm B}=2\zeta(3) \, \sigma_0 \, \left[
\ln(4\gamma^2) \,\left( l^2 - 2l + {\pi^2+12\over 12}\right)
-{2\over 3}l^3 +{30-\pi^2 \over 12}\, l - c\right] \,,
\label{47d}
\end{equation}
where
$$
c={\pi^2 +3\zeta(3)+21\over 12} = 2.873 \,, \ \ \ l=
\ln{2m_e\gamma \over \varepsilon_{\min}} \,.
$$

Using this expression for $\varepsilon_{\min} = 15 m_e$ and the results of 
a numerical integration in the region $2m_e <\varepsilon < 15 m_e$, we obtain 
the total cross section for the Born contribution:
\begin{equation}
\sigma_{\rm B}=
17.8 {\mbox{ mb for RHIC}}\,, \;\;
110 {\mbox{ mb for LHC (Pb--Pb)}}\,, \;\;
0.42 {\mbox{ mb for LHC (Ca--Ca)}}\,.
\label{47e}
\end{equation}

\subsection{An approximate calculation of the Coulomb correction}

Let us consider the correction $d\sigma_1$ defined in Eq.~(\ref{5}). Its 
contribution to the spectrum can be calculated just by the same method as it 
was done for the Born contribution. According to Ref.~\cite{GKSST} we have to 
replace in Eq.~(\ref{39})
\begin{equation}
L(x,y) -1 \to -C(\nu)
\label{48}
\end{equation}
from which it follows
\begin{equation}
d\sigma_1= -\zeta(3)\, \sigma_0\, C(\nu)\, f_1(x) \, {dx\over x}\,.
\label{49}
\end{equation}
The quantity $C(\nu)$ is defined with the help of function
$F(a,b;c;z)$ --- the Gauss hypergeometric function:
$$
C(\nu)=\int^\infty_0\, F(\tau,\nu) \, d\tau\,,\quad
z=\left({1-\tau\over 1 +\tau}\right)^2\,,
$$
\begin{equation}
F(\tau,\nu)={1\over 2\tau (1+\tau)^2}\left\{1-\left[
F(i\nu,-i\nu;1;z)\, {\pi\nu\over \sinh \pi\nu } \right]^2\right\}\,.
\label{50}
\end{equation}
The curve $C(\nu)$ and properties of the function
$F(\tau,\nu)$ are given in~\cite{GKSST}, we mention here only
that $C(\nu)=0.6734$ for Au, 0.7080 for Pb and
\begin{equation}
C(\nu)=2.9798 \, \nu^2 \;\;\mbox{ at }\;\; \nu^2 \ll 1\,.
\label{51}
\end{equation}
Analogously,
\begin{equation}
d\tilde \sigma_1=-\zeta(3) \,\sigma_0 \, C(\nu)\, f_1(\tilde x)\,
{d\tilde x \over\tilde x} \,.
\label{52}
\end{equation}

As we have explained in Sect.~\ref{select}, the spectra 
$d\sigma_1 / d\varepsilon$ and $d\tilde \sigma_1 / d \varepsilon$ have to be 
calculated with logarithmic accuracy only. Within this accuracy we have 
$x \approx \omega_1 /(2m_e \gamma)$ and 
$\tilde x \approx \omega_2 / (2m_e \gamma)$. Then for $d\sigma_1$ we can 
repeat the calculations similar to those which we have already performed to 
get the spectrum (\ref{21}) from  expression (\ref{17}). In our case this gives
\begin{equation}
d\sigma_1=
 -2\zeta(3)\,\sigma_0 \,C(\nu) \, \ln \gamma^2 \,
{d\varepsilon\over\varepsilon}\,, \quad
2m_e \ll \varepsilon \ll 2m_e\gamma\,.
\label{53b}
\end{equation}
Taking into account the same contribution from $d\tilde \sigma_1$ we find
(with logarithmic accuracy)
\begin{equation}
d(\sigma_1+\tilde\sigma_1)=
-4\zeta(3)\,\sigma_0 \,C(\nu) \, \ln \gamma^2 \,
{d\varepsilon\over\varepsilon}\,, \quad
2m_e \ll \varepsilon \ll 2m_e\gamma\,.
\label{53}
\end{equation}

The total contribution of CC to the para--Ps production is
\begin{equation}
\sigma_{\mathrm {para}}^{\mathrm {CC}}=\sigma_1+\tilde\sigma_1=-2\zeta(3) \, 
\sigma_0 \, C(\nu)\, [\ln \gamma^2 ]^2
\label{54}
\end{equation}
from which it follow that the relative contribution of CC is
\begin{equation}
{\sigma_{\mathrm {para}}^{\mathrm{CC}} \over 
\sigma_{\mathrm {para}}^{\mathrm{LLA}}} 
\approx
-{6C(\nu)\over\ln \gamma^2}\,,
\label{55}
\end{equation}
i.e. it is $-43$ \% for RHIC, $-27$ \% for LHC (Pb--Pb) and
$-2.2$ \% for LHC (Ca--Ca).

To obtain the contribution of $d\sigma_1$ to the inclusive
cross section we can explore Eq.~(\ref{50}). Strictly speaking,
the variable $\tau$ in that equation is $\tau={\bf p}^2_\perp
/(4m_e^2-q^2_1)$. But in LLA used here for $d\sigma_1$ we can
take the function $F(\tau,\nu)$ on the mass shell of the first
virtual photon (at $q^2_1=0$), i.e. we can use $\tau={\bf
p}^2_\perp /(4m_e^2)$. Now with the help of a simple relation
\begin{equation}
{dx\over x}\, d\tau={1\over 4\pi m^2_e}\,{d^3 p\over\varepsilon}
\label{57}
\end{equation}
we obtain
\begin{equation}
\varepsilon \,{d^3\sigma_1\over d^3 p}=
{\zeta(3) \over4\pi} \, {\sigma_0\over m^2_e} \, J_1\,,
\quad
J_1=-f_1(x) \,F(\tau,\nu) \,.
\label{58}
\end{equation}
Analogously, for $d\tilde\sigma_1$ we find
\begin{equation}
\tilde J_1=-f_1(\tilde x) \,F(\tau,\nu)\,.
\label{59}
\end{equation}
The function $f_1(x)$ is defined in (\ref{43}). As a result, the
inclusive cross section reads
\begin{equation}
\varepsilon \,{d^3\sigma_{\rm para}\over d^3 p}=
{\zeta(3) \over4\pi} \, {\sigma_0\over m^2_e} \, J_{\rm para}
\,,\quad J_{\rm para} =J_{\rm B}
-[f_1(x)+f_1(\tilde x)]\, F(\tau,\nu)
\label{60}
\end{equation}
where $J_{\rm B}$ is defined in~(\ref{31}), $x$ and $\tilde x$
are defined in Eqs.~(\ref{27}). In the LLA we have
\begin{equation}
f_1(x)+f_1(\tilde x) = 2 \ln{\gamma^2}\,,\;\;\;
2m_e \ll \varepsilon \ll 2 m_e \gamma \,.
\label{60a}
\end{equation}

\section{The production of ortho--Ps}

The amplitude of the ortho--Ps production can be presented in the
form (compare with  Eq.~(\ref{4}))
\begin{equation}
M_{\rm ortho}=M_1+\tilde M_1+M_2\,,
\label{61}
\end{equation}
$$
M_1=\sum_{n'=2,4,6,...} \,M_{1n'}\,,  \quad
\tilde M_1=\sum_{n=2,4,6,...} \, M_{n1}\,,\quad
M_2=\sum_{n,n'\geq 2} \,M_{nn'}\,.
$$
with sum $n+n'$ being odd. The total cross section reads (compare
with Eq.~(\ref{5}))
\begin{equation}
$$
\sigma_{\rm ortho}=\sigma_1+\tilde\sigma_1+\sigma_2\,,\quad
d\sigma_1\propto |M_1|^2\,,\quad
d\tilde\sigma_1\propto |\tilde M_1|^2\,,
\label{62}
\end{equation}
$$
d\sigma_2\propto 2{\rm Re}\,(M_1\tilde M^*_1+M_1 M^*_2 +
\tilde M_1 M_2^*)\,+ \,|M_2|^2\,.
$$
Since ${\rm Re}\,(M_1\tilde M_1^*)$  disappears after azimuthal
averaging, it is not difficult to estimate that
\begin{equation}
{\sigma_2\over \sigma_1} \sim {\nu^2\over\ln \gamma^2} < 4 \%
\label{63}
\end{equation}
for the colliders in Table~\ref{tab1}. Therefore,
\begin{equation}
\sigma_{\rm ortho}=\sigma_1+\tilde\sigma_1
\label{64}
\end{equation}
with an accuracy of the order of 4 \%; moreover, with the same accuracy we 
can calculate $\sigma_1$ and $\tilde \sigma_1$ in LLA only.

To obtain the {\bf inclusive cross section} we use the relation~(\ref{38}) 
with $dn_T$ and $dn_S$ from Eqs.~(\ref{41}) and the cross sections $d\sigma_T$ 
and $d\sigma_S$ for ortho--Ps from Eqs.~(3.22) of Ref.~\cite{GKSST}
\begin{eqnarray}
&&d\sigma_T=4\pi\zeta(3)\, {Z^4\alpha^8\over m_e^2}
\Phi^2_t(\tau,\nu)\,{d\tau\over (1+y)^3}\,,\quad
d\sigma_S=yd\sigma_T\,,
\label{65}
\\
&&\Phi_t(\tau,\nu)={1-\tau\over(1+\tau)^3}\,F(1+i\nu,
1-i\nu;2;z) \,{\pi\nu\over \sinh\pi\nu}
\nonumber
\end{eqnarray}
where $z$ is defined in Eq.~(\ref{50}).

It gives
\begin{equation}
\varepsilon \,{d^3\sigma_{\rm ortho}\over d^3
p}={\zeta(3)\over\pi} \,{\nu^2\sigma_0 \over
m_e^2} \, \Phi^2_t(\tau,\nu) \,[f_3(x)+f_3(\tilde x)]\,,
\label{66}
\end{equation}
$$
f_3(x)=\int^\infty_{x^2} \, \left(1-{x^2\over y}+y
\right) \, {dy\over y(1+y)^3} = (1+3x^2) \,\ln{1+x^2\over
x^2}\,-\,2\, -\, x^2\,{3+2x^2\over 2(1+x^2)^2} \,.
$$
The properties of the function $\Phi^2_t(\tau,\nu)$, which determines the 
angular distribution of the produced ortho--Ps, are  described in 
Ref.~\cite{GKSST}. We mention here only that the typical values of $\tau$ are 
$\tau\sim 0.1$ which correspond to a characteristic emission angle of ortho--Ps
\begin{equation}
\theta \sim {m_e\over\varepsilon} \,.
\label{67}
\end{equation}
The function $f_3(x)$ is large at small values of $x$
\begin{equation}
f_3(x)=2\left( \ln{1\over x}-1 \right) \;\; \mbox{ at } \;\;
x\ll 1
\label{68}
\end{equation}
and small at large $x$
$$
f_3(x)={1\over 2 x^4} \;\; \mbox{ at } \;\; x\gg 1\,.
$$
In the LLA we have
\begin{equation}
f_3(x)+f_3(\tilde x) \approx 2\ln \gamma^2  \,,\quad
2m_e \ll \varepsilon \ll 2m_e\gamma\,.
\label{69}
\end{equation}

For the {\bf spectrum} of ortho--Ps in the LLA we obtain the expression
\begin{equation}
d\sigma_{\rm ortho}=16\zeta(3)\, \nu^2\sigma_0 \, B(\nu)
\, \ln \gamma^2 \,
{d \varepsilon \over \varepsilon} \,, \;\;
B(\nu)=\int^\infty_0 \, \Phi^2_t(\tau,\nu) \,d\tau \,.
\label{70}
\end{equation}
Note that $B(\nu)=0.3770$ for Au, 0.3654 for Pb and
\begin{equation}
B(\nu)=0.6137-1.7073 \,\nu^2 \;\; \mbox{ at } \;\; \nu\ll 1\,.
\label{71}
\end{equation}
After integrating the spectrum (\ref{70}) over $\varepsilon$ in the region 
$2m_e \stackrel{<}{\sim} \varepsilon \stackrel{<}{\sim} 2m_e\gamma$ we obtain 
the total cross section in the LLA
\begin{equation}
\sigma_{\mathrm {ortho}}=8\zeta(3) \,\nu^2\sigma_0\,
B(\nu) \, \left[ \ln \gamma^2 \right]^2\,.
\label{72}
\end{equation}

The ortho--Ps production in  lowest order in $Z\alpha$
(production via three photons) is described by Eq.~(\ref{72}) with
$B(0)=2-2\ln 2=0.6137$. Therefore, the relative order of the
Coulomb correction (corresponding to the production with 5, 7, 9,$\dots$ 
photons)
is given by ratio
\begin{equation}
{B(\nu)-B(0)\over B(0)} \,.
\label{73}
\end{equation}
This ratio 
is equal to  $-39$ \% for the RHIC, for the LHC is reaches $-40$ \% in the
Pb--Pb mode and $-3.6$ \% in the Ca--Ca mode.

\section{Discussion}

1. We have calculated the cross sections for the production of parapositronium
(omitting terms  $\sim (Z\alpha)^2/L^2<0.4\,\%,\,L= \ln{\gamma^2}$) and 
orthopositronium (omitting contributions $\sim (Z\alpha)^2/L<4\,\%$) 
at the RHIC and the LHC. Our results are summarized in Table~\ref{tab2}.
{}From that Table we observe that the production rate is well above $10^5$ 
events per day, therefore measurements will be possible with a high precision 
at these colliders. We note that the production rate for the long living 
ortho--Ps is also relatively high, irrespectively of the suppression factor 
$(Z \alpha)^2$.

2. In these calculations we take into account not only the lowest order in the 
parameter $Z\alpha$ but the whole series in $Z\alpha$ --- the so called 
Coulomb correction. They decrease the total cross section for the para--Ps 
production by about 30 \% and 50 \% for Au--Au  and Pb--Pb collisions and by 
about 40 \% for ortho--Ps production in the same collisions. For Ca--Ca 
collisions the CC are small $\sim$ 3 \%.

3. We have presented the inclusive cross sections (see Eqs. (\ref{26}),
(\ref{31}), (\ref{60}), (\ref{66}) and
Figs.~\ref{fig2},\ref{fig3}) and the spectra for the Ps
production (see Eqs. (\ref{47}), (\ref{53}), (\ref{70}) and
Fig.~\ref{fig5}).  From these results we conclude that the number
of relativistic  Ps with a Lorentz factors up to the Lorentz
factor of the ion $\gamma$ will be large enough to be detected.

4. The signature of the events is quite clear: the colliding ions practically 
do not change their motion and after the collision they remain in their 
bunches. In the central detector no hadrons should be observed. The further 
separation of the relativistic Ps from the background can be performed using a 
method tested in a Serpukhov experiment (see Ref~\cite{Ser}). In that 
experiment a special channel with a vacuum pipe of about 40 m length was used 
and a weak transverse magnetic field inside this pipe eliminates the charged 
particles. At the end of the pipe a strong magnetic field forces the Ps to 
break into $e^+$ and $e^-$ which are easily detectable.

\acknowledgments

We are grateful to I.~Ginzburg, I.~Khriplovich and
I.~Meshkov for useful discussions. We are indebted to S.~Klein
and L.~Nemenov for clarifying the experimental situation. The
work of E.A.K.  was supported by INTAS grant 93-239 ext, the work
of G.L.K and V.G.S is supported by Volkswagen Stiftung (Az.
No.I/72 302) and by Russian Foundation for Basic Research (grant
96-02-19114).

\appendix
\section{The Born amplitude for the para--Ps production}

It is  useful to introduce the ``almost light-like'' 4-vectors
$$
\tilde P_1=P_1-{M^2\over s}P_2\,,\quad
\tilde P_2=P_2-{M^2\over s}P_1\,, \quad 
\tilde P^2_i={M^6\over s^2},\quad s=2P_1 P_2\approx
2\tilde P_1\tilde P_2 \gg M^2
$$
and to decompose the 4-vectors $q_i$
into  components in the plane of the 4-vectors $\tilde P_1$ and
$\tilde P_2$ and in the plane orthogonal to them
\begin{equation}
q_i=\alpha_i\tilde P_2+\beta_i\tilde P_1+q_{i\perp}\,,\quad
\alpha_i={2q_i\tilde P_1\over s}\,,\quad
\beta_i={2q_i \tilde P_2\over s}\,.
\label{A1}
\end{equation}
The quantities $\alpha_i$ and $\beta_i$ are the so called Sudakov
variables. It is not difficult to estimate that
\begin{equation}
{|\alpha_1|\over \beta_1} \approx {|q_1^2|\over 2q_1
P_2} < {|q^2_1|\over 4m_e M} \ll 1\,,\quad
{|\beta_2|\over \alpha_2} \approx {|q_2^2|\over 2q_2 P_1} <
{|q^2_2|\over 4m_e M} \ll 1\,.
\label{A2}
\end{equation}
In the used c.m.s. the 4-vectors $q_{i\perp}$ have $x$ and $y$
components only
$$
q_{i\perp}=(0,q_{ix},q_{iy},0)=(0,{\bf q}_{i\perp},0)\,,\quad
q^2_{i\perp}=-{\bf q}^2_{i\perp}\,.
$$

The relativistic nuclei in the Ps production can be regarded as
point-like and spinless particles. Therefore, the Born amplitude
has the form (see Fig.~\ref{fig1} for the notations)
$$
M_{\rm B}=4\pi Z^2 \alpha {(P_1+P'_1)^\mu (P_2+P'_2)^\nu \over
q^2_1 q^2_2}\; M_{\mu\nu}
$$
where $M_{\mu\nu}$ is the amplitude for the virtual process
$\gamma^* \gamma^* \to {\rm Ps}$. Due to gauge invariance we have
$$
q^\mu_1 \,M_{\mu\nu}=(\alpha_1\tilde P_2+\beta_1\tilde
P_1+q_{1\perp})^\mu \, M_{\mu \nu} =0\,.
$$ 
This expression can be
transformed to (using the estimates (\ref{A2}))
$$
(P_1+P'_1)^\mu M_{\mu\nu}=2P^\mu_1 M_{\mu\nu}
\approx -{2q^\mu_{1\perp}\over \beta_1} \,M_{\mu\nu}\,.
$$
Analogously we find
$$
(P_2+P'_2)^\nu M_{\mu\nu} \approx
-{2q^\nu_{2\perp}\over\alpha_2} \, M_{\mu\nu}
$$
which leads to the result
\begin{equation}
M_{\rm B}={16\pi Z^2\alpha\over\beta_1\alpha_2} \,{q^\mu_{1\perp}
q^\nu_{2\perp}\over q^2_1 q^2_2}\; M_{\mu\nu}\,.
\label{A3}
\end{equation}

Using the rules (2.11) from Ref.~\cite{GKSST} for the
transition from the pair production $AA\to AA e^+ e^-$ to the
process $AA\to AA+$ Ps, we obtain the Born amplitude for the
production of the para-positronium in the quantum state $n\,
^1\!S_0$
$$
M_{\rm B}={4(4\pi Z\alpha)^2 \over\beta_1\alpha_2 q^2_1
q^2_2}\, {m_e \alpha^{3/2}\over \sqrt{4\pi n^3}} \times
$$
$$
\times {1\over 4} {\rm Tr} \,\left\{ \left[\, \hat q_{1\perp}\,
{\hat p_- - \hat q_1+m_e \over (p_- -q_1)^2-m^2_e}\, \hat
q_{2\perp} \,+ \, \hat q_{2\perp}\, {\hat p_- -\hat q_2+m_e \over
(p_- -q_2)^2-m^2_e}\,\hat q_{1\perp}\right] (\hat p+m_{Ps})\,
{\rm i} \gamma^5 \right\}
$$
where $p_+=p_- = p/2$.  Taking the trace and using the
estimates (\ref{A2}) we obtain 
\begin{equation}
M_{\rm B}=-16 Z^2
\sqrt{{\pi^3\alpha^7\over n^3}} \, {s\over m_e}\, {({\bf
q}_{1\perp} \times {\bf q}_{2\perp})\cdot {\bf n}_1\over q^2_1
q^2_2} \; {4m^2_e\over 4m^2_e-q^2_1-q^2_2}\,, 
\label{A4}
\end{equation}
$$
{\bf n}_1={{\bf P}_1\over |{\bf P}_1|}\,.
$$
{}From that Born amplitude we find the inclusive cross section for
the production of para--Ps summed over quantum states $n$
\begin{equation}
\varepsilon \, {d^3\sigma_{\rm B}\over d^3 p}=
{1\over 8(2\pi)^5 s^2} \, \sum_{n=1}^{\infty} \,
\int \, | M_{\mathrm B}|^2 \,
\delta({\bf q}_{1\perp}+{\bf q}_{2\perp}- {\bf p}_\perp ) \, d^2
q_{1\perp} d^2 q_{2\perp} \,.
\label{A4a}
\end{equation}

Now we  derive a useful expression for the virtualities $q^2_i$.
{}From the obvious relations $(P_1-q_1)^2=M^2$ or
$2q_1 P_1= s \alpha_1+M^2 \beta_1= q^2_1= s\alpha_1\beta_1 +
q^2_{1\perp}$ and
$2q_2 P_1=s\beta_2 +M^2 \alpha_2= q^2_2= s\alpha_2 \beta_2+
q^2_{2\perp}$ we obtain
\begin{equation}
q^2_1={q^2_{1\perp}-(M\beta_1)^2\over 1-\beta_1},\quad
q^2_2={q^2_{2\perp}-(M\alpha_2)^2\over 1-\alpha_2}\,.
\label{A5}
\end{equation}
Further, from the relation $(q_1+q_2)^2=p^2=4m_e^2$ we get (using
the inequalities (\ref{A2}))
\begin{equation}
s\alpha_2\beta_1+ p^2_\perp = 4m_e^2 \,.
\label{A6}
\end{equation}
If we  introduce variables $x=M\beta_1/(2m_e)$ and $\tilde
x=M\alpha_2 /(2m_e)$ we immediately obtain from
Eqs.~(\ref{A5})-(\ref{A6}) and inequalities (\ref{10}) the expressions
(\ref{27})-(\ref{29}).

\section{Calculation of  $J_{\rm B}$ from Eq.~(\ref{26})}

Let us introduce the two--dimensional vectors ${\bf k}_i={\bf
q}_{i\perp}/(2m_e)$ and the quantities $V_1={\bf k}^2_1+x^2$, $V_2
={\bf k}^2_2+ \tilde x^2$, $V_3=V_1+V_2+1$. Then the integral
$J_{\rm B} $ from Eq.~(\ref{26}) transforms to
\begin{equation}
J_{\rm B}={1\over \pi}\, \int\, {({\bf k}_1\times{\bf
k}_2)^2\over (V_1 V_2 V_3)^2} \, \delta \left( {\bf k}_1+{\bf
k}_2-{{\bf p}_\perp\over 2 m_e}\right) \; d^2 k_1 d^2 k_2 \,.
\label{B1}
\end{equation}
Now we use the usual Feynman trick
$$
{1\over (V_1 V_2 V_3)^2} = 5! \, \int {\delta (\sum_j\alpha_j
-1) \prod_j \alpha_j d\alpha_j \over (\sum_j\alpha_j V_i)^6}
$$
and perform the integration over ${\bf k}_i$:
\begin{equation}
J_{\rm B}=3\tau \, \int^1_0 d\alpha_1\, \int^{1-\alpha_1}_0
d\alpha_2\,{\alpha_1\alpha_2(1-\alpha_1-\alpha_2)
(2-\alpha_1-\alpha_2)^2\over D^4} \,,
\label{B2}
\end{equation}
$$
D=(1-\alpha_1)(1-\alpha_2) \tau+ (2-\alpha_1-\alpha_2)
[1-\alpha_1 -\alpha_2+ (1-\alpha_2) x^2 +(1-\alpha_1) \tilde x^2]
$$
where $\tau$ is given in Eq.~(\ref{27}). Substituting
$$
2\alpha_1=(1+t)u,\quad 2\alpha_2=(1-t)u
$$
we present Eq.~(\ref{B2}) in the form
\begin{equation}
J_{\rm B}={3\over 8} \,\tau \, \int^1_0 du \, u^3 (1-u)
(2-u)^2  \int^1_{-1} dt { (1-t^2) \over D^4}\,,
\label{B3}
\end{equation}
$$
D=\tau \left[ 1-u+{1\over 4} u^2 (1-t^2) \right]+
(2-u) \left[ 1-u+x^2+\tilde x^2-{1\over 2} u (1-t) x^2-
{1\over 2} u (1+t) \tilde x^2 \right]
$$
which is equivalent to Eqs.~(\ref{31})-(\ref{33}).

\begin{table}
  \begin{center}
    \begin{tabular}{|l|c|c|}  \hline
      \sl Collider, mode &\sl Lorentz factor & \sl Luminosity \\
         & $\gamma=E/M$ & ${\cal L}$, ${\rm cm}^2 {\rm s}^{-1}$\\    \hline
      RHIC, Au--Au & 108 & $2\cdot 10^{26}$\\ \hline
      LHC, Pb--Pb & 3000 & $2\cdot 10^{27}$ \\    \hline
      LHC, Ca--Ca & 3750 & $4\cdot 10^{30}$ \\    \hline
    \end{tabular}
    \caption{Parameters of the discussed colliders.}
    \label{tab1}
  \end{center}
\end{table}

\begin{table}[!htb]
  \begin{center}
     \begin{tabular}{|c|c|c|c|c|}  \hline
       {\sl Collider, mode} & $\sigma_{\mathrm{para}}$  & para--Ps &
             $\sigma_{\mathrm{ortho}}$  & ortho--Ps \\
            & mb & events/sec  &  mb &  events/sec \\ \hline
           RHIC, Au--Au & 8.7 &  1.7 & 6.8 & 1.4 \\ \hline
           LHC, Pb--Pb  & 78  &  155 & 24 &  48 \\    \hline
           LHC, Ca--Ca  & 0.41&  1650& 0.0086 & 35\\    \hline
     \end{tabular}
     \caption{Cross sections and event rates at the RHIC and LHC.}
     \label{tab2}
  \end{center}
\end{table}


\end{document}